\documentclass[prl,twocolumn]{revtex4}

\usepackage{graphicx}
\usepackage{dcolumn}
\usepackage{bm}

\begin{document}

\preprint{APS/123-QED}

\title{First-principles band calculation and model construction of superconducting BiS$_2$ layers}

\author{Hidetomo Usui}
\author{Katsuhiro Suzuki}
\author{Kazuhiko Kuroki}

\affiliation{Department of Engineering Science, The University of Electro-Communications, Chofu, Tokyo 182-8585, Japan}

\date{\today}

\begin{abstract}
We construct minimal electronic models 
for a newly discovered superconductor LaO$_{1-x}$F$_x$BiS$_2$ ($T_c=$ 10.6K)
possessing BiS$_2$ layers based on first principles band calculation.
First, we obtain a model consisting of two Bi $6p$ and two S $3p$ orbitals,
which give nearly electron-hole symmetric bands. 
Further focusing on the bands that intersect the 
Fermi level, we obtain a model with two $p$ orbitals.
The two bands (per BiS$_2$ layer) 
have quasi-one-dimensional character with a double minimum dispersion, 
which gives good nesting of the Fermi surface. 
At around $x\sim 0.5$ the topology of the Fermi surface 
changes, so that the density of states at the Fermi level becomes 
large. Possible pairing states are discussed.
\end{abstract}

\pacs{74.20.-z,74.20.Pq}
\maketitle

\begin{figure}[!b]
\includegraphics[width=6cm]{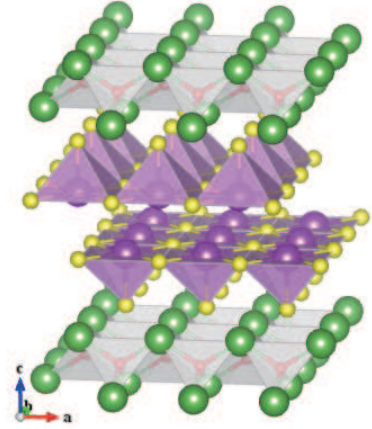}
\caption{The lattice structure of LaOBiS$_2$.
}
\label{fig1}
\end{figure}

For the past several decades, superconductors with 
layered lattice structures such as 
the cuprates\cite{Bednorz}, organic conductors\cite{organicrev},
MgB$_2$\cite{Akimitsu}, Sr$_2$RuO$_4$\cite{Maeno}, 
Na$_x$CoO$_2$\cite{Muromachi}, (Hf,Zr)NCl\cite{Yamanaka}, 
and iron pnictides\cite{Hosono} have attracted much attention. 
These layered superconductors have been of interest in many aspects 
like high $T_c$ and/or unconventional pairing mechanisms.
Quite recently, Mizuguchi {\it et al.} have discovered 
superconducting materials that possess BiS$_2$ layers, where 
the Bi and S atoms are aligned alternatively on a square lattice.
The materials found so far are 
Bi$_4$O$_4$S$_3$\cite{Mizuguchi443},
LaO$_{1-x}$F$_x$BiS$_2$\cite{Mizuguchi1112}, and 
NdO$_{1-x}$F$_x$BiS$_2$\cite{Demura}, 
which have $T_c$ of 8.6K, 10.6K, and 5.6K, respectively.
Fig.\ref{fig1} shows the lattice structure of LaOBiS$_2$, 
where partial replacement of O by F $(x=0.5)$ provides electron doping and 
gives rise to superconductivity.
These findings strongly suggest that materials with  BiS$_2$ layers 
provide yet another family for layered superconductors exhibiting 
double-digit $T_c$, and  
it is of special importance to understand the underlying electronic
structure of these materials.

Recent studies have shown that 
effective models constructed from first principles band calculation 
can provide solid basis for the study on the mechanism of superconductivity,
especially for materials with complicated band structures such as  
the iron pnictides\cite{KurokiPnictide}.
In the present paper, we perform first principles band calculation 
of LaOBiS$_2$, from which we construct 
maximally localized Wannier orbitals to obtain the effective tightbinding 
models, i.e., the  kinetic energy part of the effective Hamiltonian. 
The models consist of two or four bands that have quasi-one-dimensional 
character, which are hybridized to give a two dimensional Fermi surface.
At around $x\sim 0.5$ the topology of the Fermi surface 
changes, so that the density of states at the Fermi level becomes 
large.
We show that the quasi-one-dimensional nature results in a nesting of the 
Fermi surface, which gives rise to 
enhanced irreducible susceptibility along the diagonals that intersects 
the wave vectors $(0,0)$ and $(\pi,\pi)$. 
We discuss possible superconducting states.

\begin{figure}[!b]
\includegraphics[width=7cm]{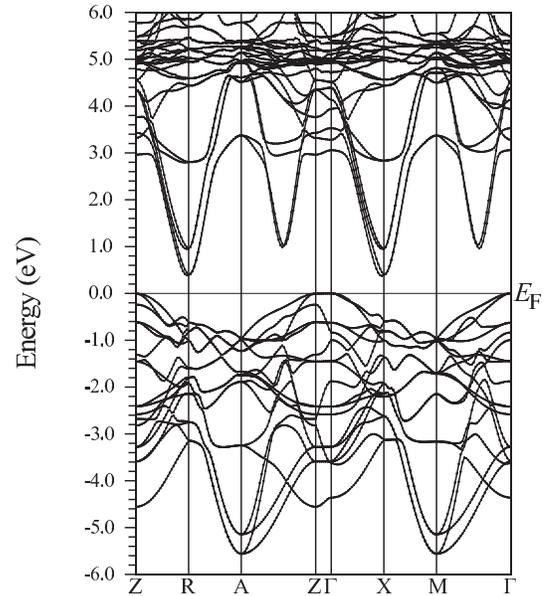}
\caption{The first principles band structure is shown.}
\label{fig2}
\end{figure}   

\begin{figure}[!b]
\includegraphics[width=7cm]{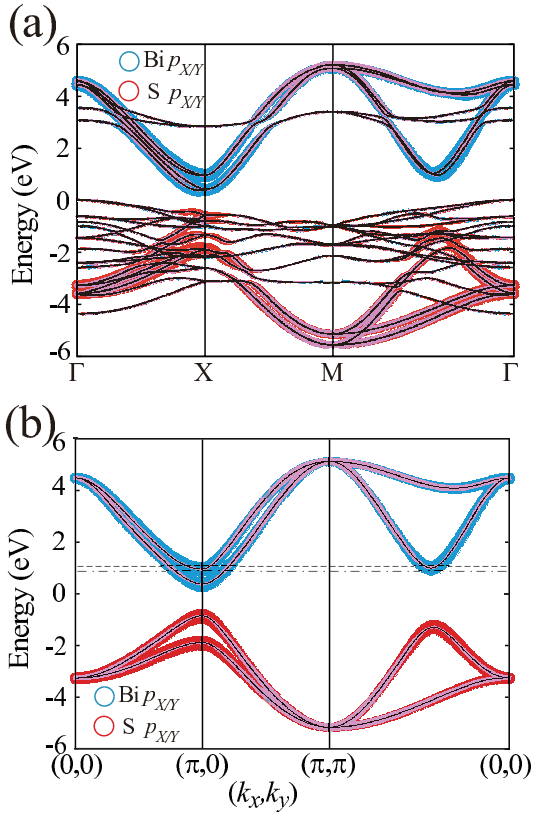}
\caption{(a) The 24 orbital model and  (b) the four orbital model.
In (b), the dashed (dotted-dashed) lines denote the Fermi energy for 
the doping ratio $\delta=0.5$ ($\delta=0.25$).}
\label{fig3}
\end{figure}

\begin{figure}[!b]
\includegraphics[width=7cm]{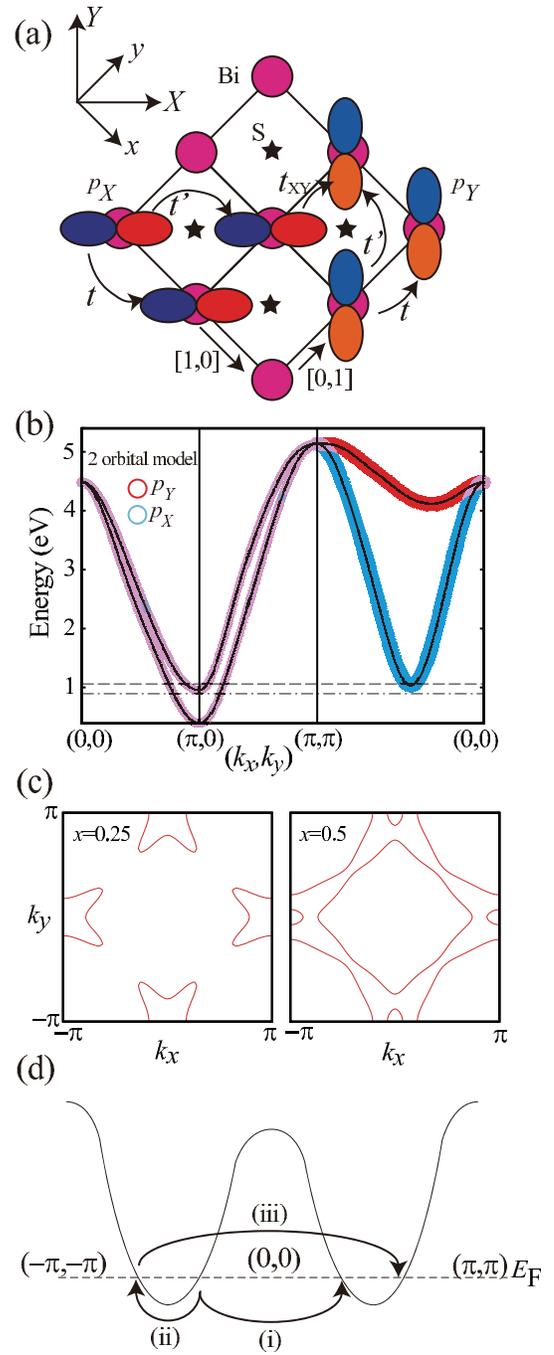}
\caption{(a) 
The tightbinding model on the Bi square lattice is shown along with 
the hoppings of $p_X$ and $p_Y$ orbitals. The stars denote the position of 
the S atom, whose $p$ orbitals are explicitly considered in the 
four orbital model, but are integrated out in the two orbital model. 
(b) The band structure of the 
two orbital model. The dashed (dotted-dashed) lines denote the Fermi energy for 
the doping ratio $\delta=0.5$ ($\delta=0.25$).
(c) The Fermi surface of the two orbital model 
for the doping rate of $\delta=0.25$ and $\delta=0.5$. 
(d) A schematic figure of the $p_X$ band along 
$(-\pi,-\pi)\rightarrow (0,0)\rightarrow (+\pi,+\pi)$. Three types of 
nesting vectors are shown by the arrows.}
\label{fig4}
\end{figure}   

The band calculation of the mother compound 
LaOBiS$_2$ is performed 
using the Wien2K package\cite{Wien2k} and adopting the 
lattice structure given in ref.\cite{structure}
(see Supplementary Material for details \cite{supp}).
Here we present results without the spin-orbit coupling, 
although this coupling does have some effect on the band structure. 
The calculation result is shown in Fig.\ref{fig2}. 
The conduction bands around the Fermi level consists mainly of  
in-plane Bi $6p$ orbital character, mixed with in-plane S $3p$.

\begin{figure}[!b]
\includegraphics[width=7cm]{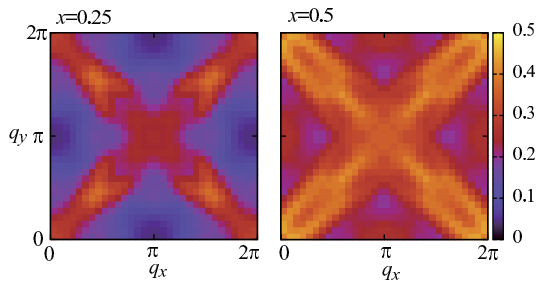}
\caption{The largest eigenvalue of the irreducible susceptibility matrix
for $\delta=0.25$ and 0.5.}
\label{fig5}
\end{figure}

\begin{figure}[!b]
\includegraphics[width=8cm]{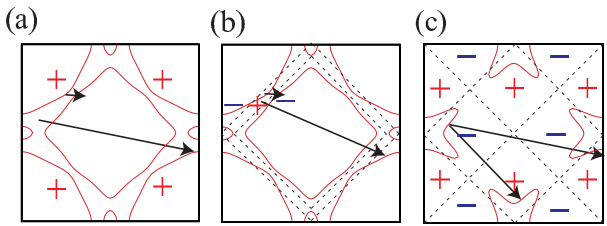}
\caption{(a) sign conserving $s$-wave 
for attractive pairing interactions such as mediated by phonons 
or charge fluctuations, 
(b) sign reversing $s$-wave, 
and (c) $d$-wave superconducting gaps 
obtained for repulsive pairing interactions mediated by spin fluctuations.}
\label{fig6}
\end{figure}   

From this band calculation, we obtain maximally localized 
Wannier orbitals\cite{Wannier,w2w}, which enables us to 
construct tightbinding models that correctly reproduces the 
original first principles band structure around the Fermi level.
First, we construct a 24 orbital model, which consists of 
six Bi $6p$, twelve S $3p$ and six O $2p$ orbitals. 
The band structure of the obtained tight binding model 
is shown in Fig.\ref{fig3}(a). Here, the thickness of the lines 
represent the weight of the in-plane $p$ orbitals within the BiS$_2$ layers.
Since the BiS$_2$ layers are without doubt 
the origin of the superconductivity, 
we can further obtain a model that omits the O $2p$ orbitals, 
the $3p$ orbitals of the out-of-plane S , and also 
the $p_z$ orbital of the in-plane S (whose bands lie away from 
the Fermi level), 
and we are left with an eight orbital model (band not shown, see 
Supplementary Material\cite{supp}).
There are eight orbitals because there are two BiS$_2$ layers per unit cell,
and each Bi and in-plane S has 
two $p$ orbitals.
Note that we should extract the portion of the bands that 
is relevant to the BiS$_2$ layers since the many body interactions (which should be included in the forthcoming studies) 
take place mainly within this layer.
By further neglecting the small 
interlayer coupling between the neighboring BiS$_2$ layers, 
we end up with  a two-dimensional 
``$p$-$p$'' four orbital model consisting of two Bi $6p$ and two S $3p$ 
orbitals. The band structure of this model is shown in Fig.\ref{fig3}(b).
We write the model Hamiltonian in the form
\begin{eqnarray}
&&H_0=\sum_{ij}\sum_{\mu\nu}\sum_\sigma
\left[t(x_i-x_j, y_i-y_j;\mu,\nu)c_{i\mu\sigma}^\dagger c_{j\nu\sigma}
\right.
\nonumber\\
&+&
\left.
t(x_j-x_i, y_j-y_i;\nu,\mu)c_{j\nu\sigma}^\dagger c_{i\mu\sigma}\right]
+\sum_{i\mu\sigma}\varepsilon_\mu n_{i\mu\sigma},
\end{eqnarray}
where $c_{i\mu\sigma}^\dagger$ creates an electron with spin $\sigma$ 
on the $\mu$-th orbital in the $i$-th unit cell, and 
$n_{i\mu\sigma}=c^\dagger_{i\mu\sigma}c_{i\mu\sigma}$. 
Then, 
the parameters of this four orbital model is given in table\ref{tab1}.
The two $p$ orbitals are denoted as $p_X$ and $p_Y$, 
where the $X$-$Y$ axis are rotated by 45 degrees from 
the $x$-$y$ axes (an figure of the four orbital model 
is not presented, but see Fig.\ref{fig4}(a) for the $X$ and $Y$ axes.) 
It is interesting to note that the upper and the lower 
bands are roughly symmetric with respect to the gap.
\begin{table}[t]
\caption{Hopping parameters 
$t(\Delta x, \Delta y; \mu, \nu)$ or the on-site energies $\varepsilon_\mu$
for the four orbital model.
The orbitals $\mu=1,2,3,4$ correspond to Bi $p_Y$, $p_X$, S
$p_Y$, $p_X$ orbitals, respectively.
We follow the notations in ref.\cite{KurokiPnictide}, i.e., 
$I$, and $\sigma_d$ corresponds to 
$t(-\Delta x, -\Delta y;\mu,\nu)$, $t(\Delta y, \Delta x;\mu,\nu)$, 
respectively, where `$\pm$' means that the corresponding  
hopping is equal to $\pm t(\Delta x, \Delta y; \mu, \nu)$,  respectively.
\label{tab1} }
\begin{tabular}{|c|ccccc|c|c|} \hline
\multicolumn{1}{|c|}{($\mu,\nu$)} & \multicolumn{5}{c}{[$\Delta x, \Delta y$]} & \multicolumn{1}{|c|}{$I$} & \multicolumn{1}{c|}{$\sigma_d$} \\\cline{2-6}
\multicolumn{1}{|c|}{} & [0,0] & [1,0] & [-1,0] & [1,-1] & [1,1] & \multicolumn{1}{c|}{} & \multicolumn{1}{c|}{} \\\hline
(1,1) & 0.890  & 0.223  & 0.223 & 0.103  & 0.082  & + & + \\
(1,2) &  & 0.100  & 0.100 &  &  & + & $-$ \\
(1,3) & 0.486  &  & -2.034  &  &  & + & + \\
(1,4) &  &  &  &  &  & + & + \\
(2,2) & 0.890  & 0.223  & 0.223 & 0.082  & 0.103  & + & + \\
(2,3) &  &  &  &  &  & + & + \\
(2,4) & 2.034  &  & -0.486  &  &  & + & + \\
(3,3) & -1.113  & -0.110  &  -0.110 & 0.100  & 0.023  & + & + \\
(3,4) &  & 0.154  & 0.154 &  &  & + & $-$ \\
(4,4) & -1.113  & -0.110  & -0.110 & 0.023  & 0.100  & + & + \\\hline
\end{tabular}
\end{table}

Even more simple model is the one which focuses only on the bands 
that intersect the Fermi level. 
By extracting these bands using the maximally localized Wannier orbitals 
centered at the Bi sites and neglecting the interlayer hoppings, 
the Hamiltonian reduces to a two dimensional 
two orbital model (Fig.\ref{fig4}(a)), whose band structure 
is shown in Fig.\ref{fig4}(b) and the hopping parameters given in 
table\ref{tab2}. Assuming a rigid band, the Fermi surface 
for this model is obtained in Fig.\ref{fig4}(c) 
for two doping ratios $\delta=0.25$ and 
$\delta=0.5$. Here $\delta$ is defined as the number of doped electrons per 
Bi site, and satisfies $\delta =x$ in an ideal situation.
At around $\delta\sim 0.5$ the topology of the Fermi surface 
changes, so that the density of states at the Fermi level becomes 
large around this band filling.
The $p_X$ and $p_Y$ bands have essentially one-dimensional 
character, where the main hopping integrals ($t'$) exist between next nearest 
neighbor sites, as shown in Fig.\ref{fig4}(a) and table\ref{tab2}. 
Thus the one-dimensional 
bands essentially have the forms 
\begin{equation}
\varepsilon_X(k)=2t'\cos(k_x+k_y),\;\;
\varepsilon_Y(k)=2t'\cos(k_x-k_y)
\label{eq1}
\end{equation}
which has a double well band dispersion along 
$(k_x,k_y)=(-\pi,-\pi)\rightarrow (0,0)\rightarrow (+\pi,+\pi)$ or 
$(-\pi,+\pi)\rightarrow (0,0)\rightarrow (+\pi,-\pi)$, 
as shown in Fig.\ref{fig4}(d).
The intra-orbital ($t$) as well as the inter-orbital ($t_{XY}$) 
nearest neighbor hoppings give the two dimensionality of the system.
\begin{table}
\caption{Hopping parameters and on-site energies for the two orbital model.
$\mu=1,2$ correspond to $p_Y$, $p_X$,  respectively. Here $t'=0.880$, 
$t=-0.167$, and $t_{XY}=0.107$}
\label{tab2}
\begin{tabular}{|c|ccccccc|c|c|} \hline
\multicolumn{1}{|c|}{($\mu,\nu$)} & \multicolumn{7}{c}{[$\Delta x, \Delta y$]} & \multicolumn{1}{|c|}{$I$} & \multicolumn{1}{c|}{$\sigma_d$} \\\cline{2-8}
\multicolumn{1}{|c|}{} & [0,0] & [1,0] & [1,-1] & [1,1] & [2,0] & [2,1] & [2,-1] & \multicolumn{1}{c|}{} & \multicolumn{1}{c|}{} \\\hline
(1,1) & 2.811  & -0.167  & 0.880  & 0.094  &   & 0.014  & 0.069  & + & + \\
(1,2) &  & 0.107  &  &  & -0.028  & 0.020  & 0.020  & + & $-$ \\
(2,1) &  & 0.107  &  &  & -0.028  & 0.020  & 0.020  & + & $-$ \\
(2,2) & 2.811  & -0.167  & 0.094  & 0.880  &  & 0.069  & 0.014  & + & + \\\hline
\end{tabular}
\end{table}

The one-dimensional nature of the bands provides good nesting of the 
Fermi surface. To see this effect, we calculate for the two orbital model  
the $4\times 4$ irreducible susceptibility matrix 
in the orbital representation, 
$\chi^{0}_{l_1,l_2,l_3,l_4}({\bf q}) 
= \sum_{\bf k}G^{0}_{l_1,l_3}G^{0}_{l_4,l_2}$,
where $G^{0}$ is the $2\times 2$ bare Green's function matrix.
In Fig.\ref{fig5}, we show the 
largest eigenvalue of the irreducible susceptibility matrix for 
the doping ratio of $\delta=0.25$ and $\delta=0.5$. The diagonal 
structures that go through (0,0) or $(\pi,\pi)$ are due to 
the nesting shown by the arrows in Fig.\ref{fig4}(d).
Note that if the nearest neighbor 
hoppings do not exist and the bands have the form in eq.(\ref{eq1}), 
the nestings (i)$\sim$ (iii) are equivalent, while this equivalence 
is lost in the presence of  $t$ (and additional hoppings).

Finally, let us discuss the possible pairing states and the 
superconducting gap structure. 
Since the many-body term of the effective Hamiltonian 
is not determined here, there are several possibilities at present.
The most relevant bands have mainly $6p$ character, which 
gives wide spread of the Wannier orbitals, 
so that the electron-electron interactions may not 
be very  strong and short-ranged as in the $3d$-orbital based materials 
such as the cuprates and iron-based superconductors. In that case, the 
electron-phonon interaction can be playing the main role  
in the Cooper pairing, and the good nesting of the Fermi surface may 
cooperate to give 
enhanced attractive pairing interaction around the nesting vectors. 
This can give rise to a 
$s$-wave pairing with constant gap sign as shown in Fig.\ref{fig6}(a). 
It is interesting to point out that Bi-based superconductors   
(Ba,K)(Bi,P)O$_3$\cite{BPBO,BKBO} are also known to 
have Fermi surface nesting, 
but with O-$2p$ and Bi-$6s$ orbital character\cite{Mattheiss}, 
differnt from the present material. 
A more exotic possibility for the same type of gap 
is related to the fact that the 
system is close to a band insulator, where electron-hole excitations might be 
playing an important role in the pairing\cite{Little}. 
In fact, if we consider electron-electron interactions in
the four orbital model, the system can be viewed as two interacting 
charge-transfer-type insulating bands. This is similar to the 
situation studied in ref.\cite{KurokiInsMetal}, where 
the occurrence of superconductivity was proposed for systems with 
a metallic band interacting with a charge-transfer-type insulating band.
In this context, it should be noted that 
the analysis for the Bi $6p$-S $3p$ four orbital model can 
give different results from those for the two orbital model 
regarding the superconducting state, 
because the band filling is different, i.e., 
the four orbital model 
is a nearly half-filled, electron-hole symmetric system , 
while the two orbital model has small band filling.

On the other hand, if we assume a short ranged repulsive interaction, 
the present two orbital model provides a 
fundamental problem of the pairing state in repulsively interacting 
quasi-one-dimensional systems, apart from its relevance to the BiS$_2$ 
layers. Namely, the occurrence of superconductivity 
in repulsively interacting 
one-dimensional systems with a double well type band 
structure was studied in the late 1990's, where 
the interaction (ii) in Fig.\ref{fig4}(d) dominates for 
certain parameter regime  
to give a gap that reverses its sign between the inner and the outer 
Fermi points\cite{Fabrizio,Kurokitrestle,Sano,KurokiNakano}. 
The existence of $t$ in Fig.\ref{fig4}(a) is essential in this case, 
because this makes the interactions 
(i)/(iii) and (ii) inequivalent.
In fact, by adding on-site intra- ($U$) and inter-orbital 
interactions ($U'=2U/3$ and the Hund's coupling and the 
pair hoppings $J=J'=U/6$), and applying 
multiorbital random phase approximation\cite{TakimotoRPA,KontaniRPA} 
to obtain the spin-fluctuation-mediated pairing interaction,  
we find the sign reversing $s$-wave gap shown in Fig.\ref{fig6}(b) 
for large doping (e.g., $\delta=0.5$ and $U=1.8$eV) and 
the $d$-wave superconducting gap shown in Fig.\ref{fig6}(c) 
for small doping (e.g., $\delta=0.25$ and $U=2.43$eV).

These gaps are determined by cooperation/competition between the 
repulsive (sign reversing) spin-singlet pairing interactions driven by the 
nestings given in Fig.\ref{fig4}(d).  The gap (b) 
indeed exhibits sign reverse between the inner and 
the outer Fermi surfaces.
On the other hand, gap (c) is obtained by reversing the gap sign 
for both interactions (i) and (ii). To make the gap have even parity 
requires additional nodes at $k_x=\pm k_y$. 
The reason why this $d$-wave is favored for small doping is because the 
diagonal nodes do not intersect the Fermi surface in this situation.
In an ideal situation, gaps (a) and (b)/(c) can be distinguished by 
$T_1^{-1}$ measurement in the nuclear magnetic resonance (NMR) experiments.
For gap (a), $T_1^{-1}$ should exhibit a coherence peak followed by a 
steep decrease, while for (b)/(c) the coherence peak 
should be (nearly) absent.
It is more difficult to distinguish (b) and (c) because 
even for the $d$-wave case (c), the nodes intersecting the 
Fermi surface are accidental.
Therefore, other phase sensitive experiments are necessary.

Finally, yet another possibility related with the quasi-one-dimensionality is 
the spin-triplet pairing.
In fact, the present band structure 
also has similarity with the quasi-one-dimensional bands 
(the hybridized $d_{xz/yz}$ bands) 
in Sr$_2$RuO$_4$, where some theoretical studies have suggested 
the possibility of spin-triplet pairing originating from these  
bands\cite{Ogata,SatoKohmoto,KurokiSRO,Takimoto}. 
NMR experiments will also provide a test for this possibility.
 
To summarize, we have obtained effective tightbinding models for 
the superconducting BiS$_2$ layers by performing 
band calculation for LaOBiS$_2$ and exploiting the 
maximally localized Wannier orbitals. 
The model consists of $p_X$ and 
$p_Y$ orbitals, and the dominant next nearest neighbor hoppings 
lead to quasi-one-dimensional bands, which give nesting  
of the Fermi surface.
Considering the quasi-one-dimensional as well as the 
doped-band-insulator nature of the band structure, 
there are several interesting possibilities of the 
superconducting state depending on the dominating 
many-body interactions.

\section{ACKNOWLEDGMENTS}

We are grateful to Y. Mizuguchi for showing us the experimental 
results for LaO$_{1-x}$F$_x$BiS$_2$ prior to publication. 
We also acknowledge all the collaborators in ref.\cite{Mizuguchi443}.
This study has been supported by 
Grants-in-Aid for Scientific Research from  MEXT of Japan and from 
the Japan Society for the Promotion of Science.


\begin{thebibliography}{99}
\bibitem{Bednorz} J.G. Bednorz and K.A. Muller, Z. Phys. B Cond. Matt. {\bf 64}
189 (1986).
\bibitem{organicrev} For a recent review, see A. Ardavan {\it et al.}, 
J. Phys. Soc. Jpn. {\bf 81}, 011004 2012.
\bibitem{Akimitsu} J. Nagamatsu {\it et al.}, Nature {\bf 410}, 63 (2001).
\bibitem{Maeno} Y. Maeno {\it et al.}, Nature {\bf 372}, 532 (1994).
\bibitem{Muromachi} K. Takada {\it et al.}, Nature {\bf 422}, 53 (2003).
\bibitem{Yamanaka} S. Yamanaka, K. Hotehama, and H. Kawaji, 
Nature {\bf 392}, 580 (1998).
\bibitem{Hosono} Y. Kamihara {\it et al.},  J. Am. Chem. Soc. {\bf 130}, 
3296 (2008).
\bibitem{Mizuguchi443} Y. Mizuguchi {\it et al.}, arXiv: 1207.3145 
\bibitem{Mizuguchi1112} Y. Mizuguchi {\it et al.}, arXiv: 1207.3558. 
\bibitem{Demura} S. Demura {\it et al.}, arXiv : 1207.5248.
\bibitem{KurokiPnictide} K. Kuroki {\it et al.}, Phys. Rev. Lett. {\bf 101}, 
087004 (2008).
\bibitem{Wien2k}
P. Blaha, K. Schwarz, G.K.H. Madsen, D. Kvasnicka, and J. Luitz, 
{\it Wien2k: An Augmented Plane Wave} + {\it Local Orbitals Program for Calculating Crystal Properties} (Vienna University of Technology, Wien, 2001).
\bibitem{structure} V.S. Tanryverdiev and O.M. Aliev, Inorg. Mater. 
{\bf 31}, 1361 (1995). 
\bibitem{supp}
See Supplemental Material for details of the first-principles calculation and the model construction.
\bibitem{Wannier} N. Marzari and D. Vanderbilt, Phys. Rev. B 
{\bf 56}, 12847 (1997); 
I. Souza, N. Marzari, and D. Vanderbilt, 
Phys. Rev. B {\bf 65}, 035109 (2001).
The Wannier functions are generated by the code developed by
A. A. Mostofi, J. R. Yates, N. Marzari, I. Souza, and D. Vanderbilt,
(http://www.wannier.org/).  
\bibitem{w2w}
J. Kunes, R. Arita, P. Wissgott, A. Toschi, H. Ikeda, and K. Held, Comp. Phys. Commun. {\bf 181} 1888 (2010).
\bibitem{BPBO} A. Wl Sleight {\it et al.}, Solid State Commun. {\bf 17}, 27 (1975).
\bibitem{BKBO} L.F. Mattheiss {\it et al.}, Phys. Rev. B {\bf 37}, 3745 (1988).
\bibitem{Mattheiss} L.F. Mattheiss and D.R. Hamann, Phys. Rev. B {\bf 28}, 4227.
\bibitem{TakimotoRPA} T. Takimoto, T. Hotta, and K. Ueda, 
Phys. Rev. B {\bf 69}, 104504 (2004).
\bibitem{KontaniRPA} K. Yada and H. Kontani, J. Phys. Soc. Jpn. {\bf 74}, 2161 
(2005).
\bibitem{Fabrizio} M. Fabrizio, Phys. Rev. B {\bf 54}, 10054 (1996).
\bibitem{Kurokitrestle} K. Kuroki, R. Arita, and H. Aoki, J. Phys. Soc. 
Jpn. {\bf 66}, 3371 (1997).
\bibitem{Sano} K. Sano {\it et al.}, J. Phys. Soc. Jpn. {\bf 74}, 2885 (2005).
\bibitem{KurokiNakano} T. Nakano, K. Kuroki, and S. Onari, 
Phys. Rev. B {\bf 76}, 014515 (2007).
\bibitem{Ogata} T. Kuwabara and M. Ogata, Phys. Rev. Lett. {\bf 85}, 4586 
(2000).
\bibitem{SatoKohmoto} M. Sato and M. Kohmoto, J. Phys. Soc. Jpn. 
{\bf 69} , 3505 (2000).
\bibitem{KurokiSRO} K. Kuroki {\it et al.}, Phys. Rev. B {\bf 63}, 060506(R) 
2001.
\bibitem{Takimoto} T. Takimoto, Phys. Rev. B {\bf 62}, 14641(R) (2000).
\bibitem{Little} W.A. Little, Phys. Rev. {\bf 134}, A1416 (1964).
\bibitem{KurokiInsMetal} K. Kuroki and H. Aoki, Phys. Rev. B {\bf 48}, 
7598 (1993).
\end{thebibliography}
\end{document}